\documentstyle[twoside,fleqn,espcrc2,psfig]{article}

\def\simlt{\stackrel{<}{{}_\sim}}
\def\simgt{\stackrel{>}{{}_\sim}}

\newcommand{\AmS}{{\protect\the\textfont2
  A\kern-.1667em\lower.5ex\hbox{M}\kern-.125emS}}

\begin{document}

\title{$Z'$ Gauge Models from Strings\thanks{Talk given at 5th International
Conference on Supersymmetries in Physics (SUSY 97), Philadephia, PA, 27-31 May
1997}
\vskip-2cm
\rightline{{ UPR--0768-T}}
\rightline{{ IEM--FT--162/97}}
\rightline{{ hep-ph/9707541}}
}

\author{Jose R. Espinosa\address{Department of Physics, University of Pennsylvania\\
Philadelphia PA 19104, USA }%
        \thanks{Work supported by the U.S. Department of Energy Grant No.
DOE-EY-76-02-3071.} }

\begin{abstract}
Potentially realistic string models often contain additional abelian
gauge factors besides the standard model group. The consequences
of such extended gauge structure are manifold both for theory and
phenomenology as I show focussing in the simplest case of just one additional
non-anomalous $U(1)'$. First, I discuss the possible symmetry breaking patterns 
according to the scale at which
the $U(1)'$ symmetry gets broken: in the first case, that scale must be below
$\sim 1$ TeV to avoid fine-tuning; in the second case, the breaking can
take place along a flat direction at an intermediate scale between the string
scale and the electroweak scale. In both cases,
I present a number of the generic implications expected, e.g. for the $\mu$
problem, $Z'$ and Higgs physics, dark matter and
fermion masses.\end{abstract}

\maketitle

\section{INTRODUCTION}

Many string models predict a gauge group at the string scale which contains,
besides the standard model group $G_{SM}=SU(3)_C\times SU(2)_L\times U(1)_Y$,
extra $U(1)$ gauge factors. In this talk, I will examine the generic
phenomenological and theoretical implications of this type of scenarios. For
simplicity, I limit
the analysis to one extra non-anomalous $U(1)'$. This is the simplest gauge
extension of the SM one can consider and the phenomenology of particular $Z'$
models has been studied extensively. Here, no attempt is made at building a
concrete realistic string model but rather at exploring the generic implications
of string inspired $Z'$ models. As we shall see, such simple extended models 
can have manifold implications for, among others, the $\mu$-problem, Higgs
physics, exotic matter, fermion masses, dark matter, the spectrum of
superpartners and cosmology.
I will describe some of these implications in two different scenarios which can be
distinguished by the scale of $U(1)'$ breaking. For more details the reader is
directed to refs.~\cite{cl,cdeel,cceel,clrev}.

Let me be more specific about the string input in the analysis. I will consider a
class of specific string vacua: orbifold models \cite{dhvw,fiqs} with Wilson
lines, in particular,
models based on the free-fermionic construction (at special points of moduli
space) \cite{abk,nahe,fara,chl}. I then assume:
\begin{itemize}
\item N=1 Supersymmetry
\item The (visible sector) gauge group at $M_{string} (\simeq 5\times 10^{17}\
GeV$) is $G_{SM}\times U(1)'$,
where the extra abelian factor has no anomaly.
\item The matter content is that of the Minimal Supersymmetric Standard Model
(MSSM) plus some number of SM singlets, $S_i$, which carry $U(1)'$ charges $Q_i$,
plus some exotics (generally transforming non-trivially under $G_{SM}$). Both
singlets and exotics appear generically in string models and can play an important
role in breaking the $U(1)'$ symmetry. 
\item The gauge couplings $g_3,g_2,g_1,g'_1$ (for $SU(3)_C$, $SU(2)_L$, $U(1)_Y$
and $U(1)'$ respectively, with proper normalization for $g_1$ and $g_1'$
\cite{cdeel}) unify at $M_{string}$. This unification works only at the $10-15\%$
level in the simplest models (in which the couplings meet at a scale below
$M_{string}$). The small numerical inconsistency that results does not
affect the generic predictions of these models that I will present.
\item Supersymmetry breaking, the main uncertainty in extracting phenomenological
consequences from string models, is parameterized by soft masses for scalars and
gauginos and by trilinear masses associated to superpotential terms. Naturalness
requires these masses (which I denote in general by $m_{soft}$) to be of the
order of the electroweak scale. 
\item No bilinear mass terms are present in the superpotential and the Yukawa
couplings at $M_{string}$ are either zero
or related to the unified gauge coupling $g_0$ by $\sqrt{2}$ factors (typically
$\sim g_0\sqrt{2}, g_0$ or $g_0/\sqrt{2}$ in the fermionic $Z_2\times Z_2$
orbifold constructions).
\end{itemize}

For fixed values of the parameters at the string scale one can address, by a
renormalization group analysis (with the chosen parameters as boundary conditions)
what are the low-energy implications typical of these scenarios.
 
\section{RADIATIVE BREAKING OF $U(1)'$}

To have an acceptable model, the extra $U(1)'$ gauge factor has to be broken at
low energies. Generically, this breaking is achieved radiatively (in a way which
parallels electroweak radiative breaking in the MSSM \cite{exact}) when the soft
mass squared $m_S^2$ of
some singlet $S$ [charged under $U(1)'$] is driven to negative values in the
infrared
(an alternate breaking mechanism is discussed in the next section).
This causes an instability of the potential for $\langle S\rangle=0$, and $S$
acquires a non-zero vacuum expectation value (VEV) thus breaking the $U(1)'$. The
value of $\langle S\rangle$,
and thus the breaking scale, is fixed by quartic $D$ terms [from $U(1)'$] in the
potential for the field $S$ (alternatively there may be also quartic $F$-terms),
\begin{equation}
V(S)=m_S^2S^2+\frac{1}{2}{g_1'}^2Q_S^2S^4,
\end{equation}
giving $g'\langle S\rangle\sim |m_S|$.

If the two Higgs doublets $H_1$, $H_2$ are charged under $U(1)'$ (the interesting
case) electroweak breaking is influenced at tree-level by the breaking of $U(1)'$.
In order to have a natural electroweak breaking scale (or equivalently, to avoid
fine tuning the soft masses), the $U(1)'$ breaking scale cannot be much larger
than
${\cal O}(1)\ TeV$. Thus, we see that the $Z'$ models derived from strings
naturally predict $M_{Z'}\simlt {\cal O}(1)\ TeV$ \cite{cl}.

One exception to this generic result [besides the trivial one of having the
$U(1)'$ decoupled from the electroweak Higgs sector] occurs if the $U(1)'$
is broken radiatively along a flat direction. For example, if the model contains
two
SM singlets $S_{1,2}$ with opposite $U(1)'$ charges, the field direction $S\equiv
S_1=S_2$ is $D$-flat and (assuming also $F$-flatness) the tree-level potential
$V(S)=-m_S^2S^2$ has no quartic coupling and is unstable. Eventually, the
potential gets stabilized at very large values of the field $S$ due to radiative
corrections [the most important of which are encoded in the running mass
$m_S^2(S)$] or by non-renormalizable operators (NROs) which can no longer be
neglected for such large field values. In both cases the 
$U(1)'$ breaking scale is much larger than the electroweak scale (e.g. $10^{12}\
GeV$). This is one example of an intermediate scale being generated in string
models \cite{fabio}.
Still, these scenarios have also many implications for the low-energy
effective theory. In the rest of the talk I will discuss separately the two
general classes of scenarios just described: I) $Z'$ below the $TeV$ scale
\cite{cdeel} and II) intermediate scale $Z'$ \cite{cceel}.  For other related
discussions in the literature see \cite{rela}.

\section{$Z'$ BELOW THE TeV SCALE}

Let us consider the simplest case, which requires the presence of just one singlet
$S$ to give a correct symmetry breaking pattern. We further assume that the
$U(1)'$ charge asignments of $S$ and the two Higgs doublets $H_{1,2}$ are such
that the superpotential contains a term
\begin{equation}
\label{shh}
W =h_s S H_1\cdot H_2,
\end{equation}
and thus $Q_S+Q_1+Q_2=0$.
After symmetry breaking, $\langle S\rangle=s/\sqrt{2}$, $\langle H_1^0\rangle=
v_1/\sqrt{2}$ and $\langle H_2^0\rangle=v_2/\sqrt{2}$, the gauge group gets
broken down to $U(1)_{em}$ with 
\begin{equation}
M_W^2=\frac{1}{4}g_2^2(v_1^2+v_2^2),
\end{equation}
from which $v^2\equiv v_1^2+v_2^2 = (246\ GeV)^2$, and
\begin{eqnarray}
\label{Zmatrix}
M^{2}_{Z-Z'}=\left (\begin{array}{c c} 
M_{Z}^{2}&\Delta^{2}\\\Delta^{2}&M_{Z'}^{2}\end{array}\right),
\end{eqnarray}
where 
\begin{eqnarray}
M_{Z}^{2}&=&\frac{1}{4}G^2(v_{1}^2+v_{2}^2),\\
M_{Z'}^{2}&=&{g'}_{1}^{2}(v_{1}^{2}Q_{1}^{2}+v_{2}^{2}Q_{2}^{2}+s^{2}Q_{S}^{2}),\\
\label{mix}
\Delta^{2}&=&\frac{1}{2}g_{1}'\,G(v_{1}^2Q_{1}-v_{2}^2Q_{2}).
\end{eqnarray} 
Here we use $G^2=g_2^2+g_Y^2=g_2^2+3g_1^2/5$.

From Eqs.~(\ref{Zmatrix}) and (\ref{mix}) we see that it is possible to 
accomodate a very small $Z-Z'$ mixing angle (the experimental upper bound is
${\cal O}(10^{-3})$ in typical models \cite{presentz}) due to a cancellation in
the off-diagonal term $\Delta^{2}$ for a particular value of $\tan\beta\equiv
v_2/v_1\sim\sqrt{|Q_1/Q_2|}$.
The presence of the singlet VEV $s$, on the other hand, helps in giving to $Z'$ 
a sufficiently large mass to evade collider constraints \cite{cdflim}.
In ref.~\cite{cdeel} two phenomenologically viable scenarios were identified:
\begin{itemize}
\item {\bf (a) $A$-driven scenario.} Symmetry breaking is driven by a large value
of the trilinear soft mass associated to the superpotential term (\ref{shh})
[this is similar to $SU(3)_C$ color breaking in the MSSM for
large values of $A_Q$, the trilinear coupling associated to the quark Yukawa
couplings],
that is, $A\gg M_{1/2}, m_0$, where $M_{1/2}$ represents a typical gaugino mass
and 
$m_0$ a scalar soft mass. In this scenario, the minimum of the potential has
$v_1\simeq
v_2\simeq s$. This is shown in figure~1, where the adimensional ratio
$f_i=v_i/m_0$
is plotted as a function of $c_A=A/m_0$ in some particular example. For $Q_1=Q_2$,
values of $\tan\beta$
close to 1 would give a small enough mixing angle.

\item {\bf (b) Large $s$ scenario. ($s\gg v_1,v_2$)} This case corresponds to a
hierarchy in the masses $M_{Z'}\gg M_Z$. The electroweak scale, as represented by
$M_Z$, is light due to
accidental cancellations among soft masses, which are typically of the order
$M_{Z'}$. To avoid fine-tuning, $M_{Z'}$ cannot be much larger than ${\cal O} (1\
TeV$) as we already discussed before.
\end{itemize}

\begin{figure}[hbt]
\centerline{
\psfig{figure=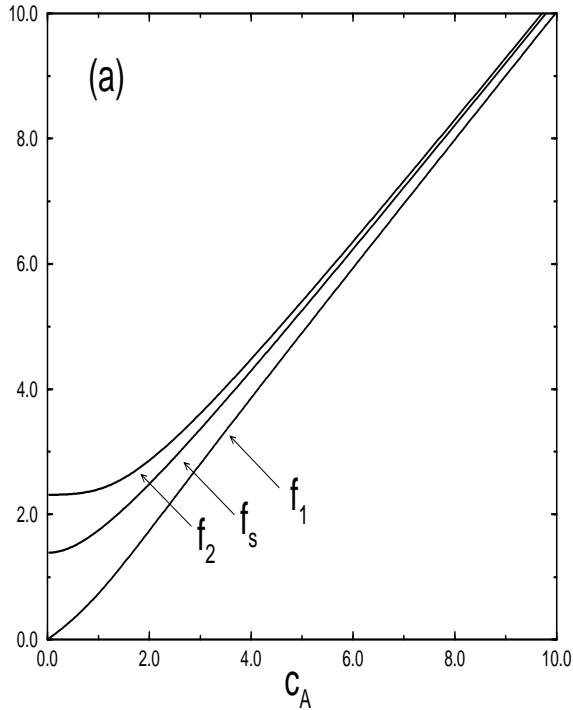,height=9cm,width=8cm,angle=-90,bbllx=3.cm,bblly=3.cm,bburx=19.cm,bbury=24.cm}}
\caption{\footnotesize
 $c_{A}$ dependence of the dimensionless field VEVs for the $A$-driven scenario
with $h_s=0.7$ and $Q_1=Q_2=-1$.}
\end{figure}

\subsection{Soft Masses at $M_{string}$}

One can then proceed to determine what pattern of soft supersymmetry breaking
terms at the string scale will lead, via RG evolution, to the two low-energy 
acceptable scenarios described previously.

With the minimal particle content (MSSM + singlet) it is not possible to achieve
the correct symmetry breaking if universal boundary conditions for the soft masses
at $M_{string}$ are used. Both
scenarios, (a) and (b), can however be reached with non-universal boundary
conditions. The allowed parameter space is somewhat constrained
and typically corresponds to gaugino masses much smaller than the scalar masses.

It is easier to have an acceptable symmetry breaking pattern in the presence of
exotics, in particular if the singlet $S$ has couplings to them of the form
\begin{equation}
W=h_E SE_1E_2.
\end{equation}
The large value of the Yukawa coupling $h_E$ ($\sim g_0$) triggers easily $U(1)'$
breaking and the large $s$ scenario [(b)] can be reached even with universal soft
masses. The large $A$ scenario [(a)] is more constrained and some degree of
nonuniversality is required.

\subsection{The $\mu$-problem}

In the presence of the superpotential term $h_sSH_1\cdot H_2$, the $\mu$ term
of the MSSM Higgs potential is generated dynamically by the VEV of the singlet $S$
as
\begin{equation}
\mu_{eff}=h_s\langle S\rangle.
\end{equation}

As discussed before, $\langle S \rangle$ lies at the electroweak scale, so that
$\mu_{eff}$ is of the correct order of magnitude to give a proper electroweak
breaking [the same mechanism generates a
$B$ term $\sim h_s\langle S\rangle A$$\sim m_{soft}$]. This solves the $\mu$
problem
\cite{muprob}
of the MSSM (where $\mu$ is a SUSY parameter which does not know about the
electroweak scale and has to be set of that order by hand).

This solution to the $\mu$ problem is similar in spirit to the mechanism in the
Next to Minimal Supersymmetric Standard Model (NMSSM) \cite{nmssm} in which the
MSSM is
extended by a chiral singlet but no additional $U(1)$ gauge symmetry is present.
The advantage of having a $U(1)'$ present is that there is no cosmological problem
with the formation of domain walls after symmetry breaking: the discrete $Z_3$
symmetry of the NMSSM responsible for the domains is embedded in $U(1)'$.

This solution is complementary to the Giudice-Masiero mechanism \cite{gm},
in the sense that the $U(1)'$ symmetry would forbid a $H_1H_2$ term in the
K\"ahler potential, which is in the basis of that mechanism.

The two scenarios (a) and (b) would correspond to a small and large $\mu_{eff}$
respectively. In the first scenario ($A$ dominated) light charginos and
neutralinos are expected, with obvious implications for LEP~II searches
\cite{lep}.

\subsection{Higgs Physics}

The (CP conserving) Higgs sector contains two doublets $H_1$, $H_2$ and one
singlet $S$ making up a total of 10 degrees of freedom. After spontaneous symmetry
breaking, 4 d.o.f. are eaten by the longitudinal components of $Z$, $Z'$ and
$W^\pm$. The 6 d.o.f. left appear as physical Higgs bosons: three (CP even)
scalars $H_i^0$, one (CP odd) pseudoscalar $A^0$ and one charged pair $H^\pm$.
That is, the model contains just one scalar more than in the MSSM.

The pattern of Higgs masses is very different in the two scenarios (a) and (b):

In the $A$-dominated scenario, all Higgs masses are controlled by one single mass
scale $v$ and three adimensional parameters: $h_s$, $g_1'$ and $Q_1$. The spectrum
is therefore
light, although all Higgses can still evade detection at LEP~II. One typical
example, for $h_s=0.7$, ${g_1'}^2=(5/3)G^2\sin^2\theta_W$ and $Q_1=-1$, has
$m_{H^\pm}=146\ GeV$, $m_{A^0}=211\ GeV$, $m_{H_3^0}=230\ GeV$,
$m_{H_2^0}=152\ GeV$ and $m_{H_1^0}=122\ GeV$ (tree-level masses).

In the large $s$ scenario there are three mass scales in the Higgs spectrum.
Typically $m_{H_3^0}\simeq M_{Z'}$ (with $H_3^0$ singlet dominated),
$m_{A^0}\gg M_Z$, with $(H_2^0,A^0;H^\pm)$ forming a nearly degenerate doublet
decoupled from electroweak symmetry breaking, and $H_1^0$ remains at the
electroweak scale.

An upper bound on the tree-level mass of $H_1^0$ can be easily obtained as
\begin{equation}
\label{bound} 
m_{H_1^0}^2\leq M_Z^2\cos^22\beta+\frac{1}{2}h_s^2v^2\sin^2
2\beta+{g'}_1^2{\overline Q}_H^2v^2,
\end{equation}
with ${\overline Q}_H=Q_1\cos^2\beta+Q_2\sin^2\beta$. The first term  in
Eq.~(\ref{bound}), coming from $SU(2)_L\times U(1)_Y$ $D$-terms, is the MSSM
piece. The second, from $F$-terms, appears also in the NMSSM but the third piece,
from $U(1)'$ $D$-terms, is a particular feature of this type of models 
\cite{habsh,denis,xg}. More refined bounds can be obtained under extra assumptions
for the parameters of
the model (see \cite{cdeel}). This tree-level bound can be very large; $170\ GeV$
is a typical example (it is usually forgotten that the bound $\sim 150\ GeV$
for any perturbative SUSY model \cite{xg} applies only if the gauge sector is not
extended) so that $H_1^0$ can easily evade detection at LEP~II.

\subsection{Superpartner Spectrum}

The spectrum of superpartners is also typically light for the $A$-dominated
scenario (with all particles expected to lie at the electroweak breaking scale)
while in the large $s$ scenario soft masses are generically larger, giving a
decoupled SUSY mass pattern.

There is, however, an extra effect associated with $U(1)'$ $D$-term contributions
to the masses. The squared mass of any scalar $\phi_i$ (with $U(1)'$ charge $Q_i$)
is shifted by the amount
\begin{equation}
\label{sqdterm}
\delta m_i^2 = \frac{1}{2}{g'}_1^2Q_i(Q_1v_1^2+Q_2v_2^2+Q_Ss^2).
\end{equation}
This can be a sizable effect that can cause some superpartner to be lighter. In
general, the mass sum-rules that can be derived in particular models will be
violated by these contributions, leaving a characteristic imprint in the mass
spectrum \cite{drees}.

\subsection{Dark Matter}

The neutralino sector in these models is composed of
$\tilde{B}$, $\tilde{W_3}$, $\tilde{H}_1^0$, $\tilde{H}_2^0$ plus
$\tilde{B}'$ and $\tilde{S}$.
In most of the parameter space the LSP (which is the natural candidate for dark
matter when $R$-parity is conserved) is the lightest neutralino with composition
dominated by $\tilde{B}$. However, in some regions of parameter space (with large
$s$ and large gaugino masses) the lightest neutralino is $\tilde{S}$-dominated
making an atypical dark matter candidate \cite{dce} with interesting  
rates for direct detection ($\sim 10^{-2}$ events/kg/day in $^{73}Ge$ detectors).

\section{INTERMEDIATE SCALE $Z'$}
\begin{figure}[hbt]
\centerline{
\psfig{figure=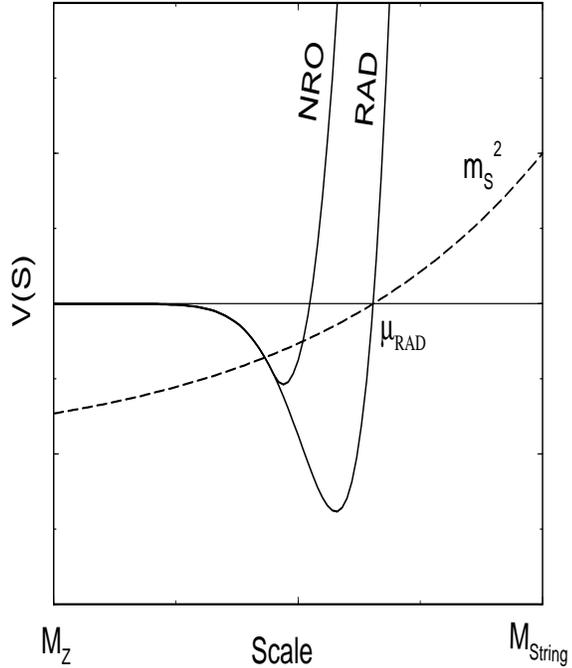,height=9cm,width=8cm,angle=-90,bbllx=3.cm,bblly=3.cm,bburx=19.cm,bbury=24.cm}}
\caption{\footnotesize
Effective potential (solid lines) along the flat direction: {\bf 1.} (marked
$RAD$) in
the case of stabilization by the running of $m_S^2$ (dashed line) and {\bf 2.}
(marked $NRO$) in the case of stabilization by non-renormalizable operators below 
the scale $\mu_{RAD}$.} \end{figure}

In this case, at least two SM singlets, $S_1, S_2$, with $Q_1Q_2<0$ are required.
Calling $S$ the flat direction determined by the condition $|Q_1|S_1^2=|Q_2|S_2^2$
(we assume
also $F$-flatness along $S$) the renormalization-group improved potential for $S$
is simply
\begin{equation}
V(S)=m_S^2(S)S^2.
\end{equation}
If the couplings of $S$ are such that $m_S^2<0$ in the infrared (we assume
$m_S^2>0$ at the string scale) the shape of the potential is as depicted in
figure~2 (solid line labeled RAD). The potential tracks the running of $m_S^2$ 
(dashed line) and is
stabilized when $m^2_S$ turns positive at the scale $\mu_{RAD}$. The minimum of
the potential occurs at some intermediate scale $m_I\simlt \mu_{RAD}\gg M_Z$.

If $\mu_{RAD}$ is very large, non-renormalizable operators like 
\begin{equation}
\label{wnro}
W_{NRO}\sim\alpha_K\left(\frac{S}{M}\right)^KS^3,
\end{equation}
which are generically present after integrating out physics at the string scale
($M\simeq M_{string}$) cannot be neglected in the determination of the
minimum. In (\ref{wnro}), we write
the NRO term, originally given by powers of $S_1$ and/or $S_2$, projected
along the flat direction $S$.  These terms
contribute to the
potential a positive-definite term 
\begin{equation}
\delta V(S)\sim \left(\frac{S^{2+K}}{M^K}\right)^2,
\end{equation}
that lifts the flat direction and can stabilize the potential below $\mu_{RAD}$.
The shape of the potential in this case is represented in figure~2 by the solid
line labeled NRO.

We can distiguish then
two different classes of models according to the mechanism for stabilization of
the potential:
\begin{itemize}
\item {\bf 1.} Stabilization by pure running of $m_S^2$ (typical case for
$\mu_{RAD}\simlt
10^{12}\ GeV$). The minimum is fixed at $\langle S\rangle\sim \mu_{RAD}$.

\item {\bf 2.} Stabilization by NROs (typical case for $\mu_{RAD}\simgt 10^{12}\
GeV$). The minimum is fixed at
\begin{equation}
\label{muk}
\langle S\rangle\sim \mu_K\equiv (m_{soft}M^K)^\frac{1}{K+1}< \mu_{RAD}.
\end{equation}
\end{itemize}

\subsection{Radiative Breaking}

Both scenarios, 1 and 2, are based on radiative breaking of the $U(1)'$ symmetry
along the flat direction $S$.
The precise condition that has to be satisfied at the electroweak scale to ensure
the breaking (the $m_S^2<0$ condition of the previous subsection) is
\begin{equation}
\label{flatbreak}
\frac{m_1^2}{|Q_1|}+\frac{m_2^2}{|Q_2|}<0,
\end{equation}
where $m_1^2, m_2^2$ are the (electroweak scale) soft masses of $S_1$ and $S_2$
respectively. From Eq.~(\ref{flatbreak}) we see that $m_1^2<0$ and/or $m_2^2<0$
are required [and a hierarchy in the charges can make (\ref{flatbreak}) easier to
fulfill].

This breaking can be easily achieved when $S_1$ (and/or $S_2$) couples to exotics
in a superpotential term
\begin{equation}
W=h_ES_iE_1E_2.
\end{equation}
In \cite{cceel} different cases (with different quantum numbers and multiplicities
for the exotic fields $E_i$) were studied. It was found that a correct breaking
can be obtained even in the case of universal soft masses at the string scale 
(provided the multiplicity of the exotics is large enough). By varying the
parameters of the model the scale $\mu_{RAD}$ can be moved in the wide range
$10^5\ GeV$ to $10^{16}\ GeV$.

\subsection{Low-Energy Effective Theory}

In both cases, 1 and 2, the $U(1)'$ breaking takes place at an intermediate
scale (e.g. $10^{12}\ GeV$) much larger than the electroweak scale. The $Z'$
boson (and the exotics coupled to the flat direction) is thus very heavy (this can 
have interesting effects for gauge coupling unification \cite{bl}). The
effective theory at the TeV scale is just the MSSM, but with an extra chiral
multiplet $\hat{S}$, remnant of $U(1)'$ breaking. However, the interactions of
$\hat{S}$ with MSSM fields are suppressed by powers of the intermediate scale.
Nevertheless, these scenarios are interesting because many parameters of the MSSM
can be determined by intermediate scale physics. 

For example, the mass spectrum of superpartners is affected by the intermediate
scale $U(1)'$ breaking. 
At the renormalizable level, the only interaction between the MSSM fields
and the intermediate scale fields arises from the $U(1)'$ $D$-terms in the scalar
potential. The resulting effect after integrating out the fields which have heavy
intermediate scale masses \cite{drees} is a shift of the soft masses of MSSM
fields charged under the extra $U(1)'$:  
\begin{equation} 
\delta
m_i^2=-Q_i\frac{m_1^2-m_2^2}{Q_1-Q_2}.  
\end{equation}

\subsection{The $\mu$ parameter}

In these models, the $\mu$ parameter can be generated dynamically by the
VEV of $S$ through non-renormalizable operators in the superpotential that couple
the MSSM Higgs doublets to the flat direction \cite{muprob}:
\begin{equation}
\label{wmuflat}
W=h S H_1\cdot H_2 \left(\frac{S}{M}\right)^P.
\end{equation}
From this term, the following effective $\mu$ parameter results
\begin{equation}
\label{muflat}
\mu_{eff}=h\frac{\langle S\rangle^{P+1}}{M^P}.
\end{equation}
Its order of magnitude in both scenarios, 1 and 2, is:
\begin{itemize}
\item {\bf 1. Radiative Stabilization.} In this case $\langle S\rangle\sim
\mu_{RAD}$. Depending on the value of $P$ in Eq.~(\ref{muflat}), the value of
$\mu_{RAD}$ must be tuned to give $\mu_{eff}\sim m_{soft}$:
\begin{equation}
\label{radtun}
\mu_{RAD}\sim (m_{soft}M^P)^{\frac{1}{P+1}}.
\end{equation} 

\item {\bf 2. Stabilization by NROs.} Eq.~(\ref{radtun}) is exactly of the form
expected for $\mu_K$ [see Eq.~(\ref{muk})], the scale of the minimum in these
scenarios. In other words, for $K$ in Eq.~(\ref{wnro}) equal to $P$ in
Eq.~(\ref{wmuflat}),
one gets automatically $\mu_{eff}\sim m_{soft}$.
The same applies to the $B$-term in the MSSM Higgs potential, which can be
generated by $\langle S\rangle$ in a similar way.
\end{itemize}

\subsection{Non-Renormalizable Operators}

As we have seen, NROs play a crucial role in these intermediate scale models.
Interestingly, NROs are
calculable in a large class of string models \cite{kalara,faraggi96a}.
Writing 
\begin{equation}
W=\alpha_K\left(\frac{S}{M_{Pl}}\right)^K\equiv\left(\frac{S}{M_{K}}\right)^K,
\end{equation}
the coefficient $\alpha_K$ is basically given by a complicated world-sheet
integral and some Clebsch-Gordan coefficients. Reabsorbing $\alpha_K$ in the mass
$M_K$, the general tendency is that $M_K$ increases with $K$ so that the weight
of higher order terms is further suppressed \cite{kalara,faraggi96a,cceel}.

It is also important to determine which NROs will be present, that is, what values
of the powers $P$ (for the $\mu$ parameter) and  $K$ (to fix the minimum) will be
allowed. The values of these powers are constrained by
\begin{itemize}
\item $U(1)'$ gauge invariance. For example, if $Q_1=4/5$ and $Q_2=-1/5$,
\begin{equation}
W\sim (S_1S_2^4)^n\sim S^{5n},
\end{equation}
so that $K=5n$.
\item World-sheet selection rules, e.g. in free-fermionic models, can forbid
terms otherwise consistent with all gauge symmetries (see, for example,
\cite{faraggi96a,rizos91}). So, one should keep in mind that not all the
NROs satisfying all gauge symmetry constraints will be necessarily present in the
superpotential. This is a typical stringy feature which is in contrast with the
field theory lore.
\end{itemize}

\subsection{Fermion Masses}

The large value of string Yukawa couplings offers a natural explanation
for the large value of the top mass \cite{faraggi}. The light fermions would have
no
Yukawa couplings in the superpotential and would therefore be massless,
which is a good first order approximation.
Models with an intermediate scale provide a mechanism to give small non-zero
masses to these fermions using NROs like
\begin{eqnarray}
W_{u_i}&\sim & {H}_2\cdot {Q}_i {U}^c_i
            \left( {{S}\over M} \right)^{P^{'}_{u_i}},\nonumber\\
W_{d_i}&\sim & {H}_1\cdot {Q}_i {D}^c_i
            \left( {{S}\over M} \right)^{P^{'}_{d_i}},\\
W_{e_i}&\sim &{H}_1\cdot {L}_i {E}^c_i
            \left( {{S}\over M} \right)^{P^{'}_{e_i}},\nonumber
\label{fermas}
\end{eqnarray}     
where $i$ is a generation index and I use standard notation for the
MSSM chiral supermultiplets.
Small Yukawa couplings are generated after $U(1)'$ symmetry breaking:
\begin{equation}
\label{yuk}
Y\sim\left(\frac{\langle S\rangle}{M}\right)^{P'}.
\end{equation}
Limiting the discussion to scenarios with NRO stabilization, which give more
definite predictions, it was shown in \cite{cceel} how the family pattern
of quark and lepton masses \cite{barnett96} can be qualitatively reproduced for
particular choices of the
powers $P'$ in Eq.~(\ref{fermas}) and $K$ in Eq.~(\ref{wnro}). In particular,
as shown in Table~1, $K=5,6$ in combination with $P'=1$ for the second family and
$P'=2$ for the first
are satisfactorily close to the observed pattern. The mass of the light fermions
of the third family do not fit quite as well and call for some other mechanism.
However, given the roughness of the estimates and the simplicity of the model, the
overall pattern of masses is quite encouraging.
\begin{table*}[hbt]
\caption{Non-Renormalizable
MSSM mass terms via $\langle S \rangle$. For
$m_{soft}\sim 100$ GeV, $M\sim 3\times 10^{17}$ GeV.}
\label{masstable2}
\begin{tabular*}{\textwidth}{cccccccc}
\hline
\hline
       ~& $P$ or $P'$ & $K=2$ & $K=3$ & $K=4$ & $K=5$
                                      & $K=6$ & $K=7$\\
\hline
\hline
$\left(\frac{m_{soft}}{M}\right)^{\frac{1}{K+1}}$
&
& $7\times 10^{-6}$
& $1\times 10^{-4}$
& $8\times 10^{-4}$
& $3\times 10^{-3}$
& $6\times 10^{-3}$
& $1\times 10^{-2}$
      \\
\hline
$\langle S\rangle$ (GeV)
& 
& $2\times 10^{12}$
& $4\times 10^{13}$
& $2\times 10^{14}$
& $8\times 10^{14}$
& $2\times 10^{15}$
& $3\times 10^{15}$
      \\
\hline
& $K-1$
& $1\times 10^{5}$
& $7\times 10^{3}$
& $1\times 10^{3}$
& $400$
& $200$
& $90$
      \\
$\frac{\mu_{eff}}{m_{soft}}$ 
& $K$
& 1
& 1
& 1
& 1
& 1
& 1
      \\ 
& $K+1$
& $7\times 10^{-6}$
& $1\times 10^{-4}$
& $8\times 10^{-4}$
& $3\times 10^{-3}$
& $6\times 10^{-3}$
& $1\times 10^{-2}$
      \\
\hline
& $0$
& 1
& 1
& 1
& 1
& 1
& 1
      \\ 
& $1$
& $7\times 10^{-6}$
& $1\times 10^{-4}$
& $8\times 10^{-4}$
& $3\times 10^{-3}$
& $6\times 10^{-3}$
& $1\times 10^{-2}$
      \\
& $2$
& $5\times 10^{-11}$
& $2\times 10^{-8}$
& $6\times 10^{-7}$
& $7\times 10^{-6}$
& $4\times 10^{-5}$
& $1\times 10^{-4}$
      \\
$\frac{m_{Q,L}}{\langle H_i\rangle} $ 
& $3$
& $3\times 10^{-16}$
& $2\times 10^{-12}$
& $5\times 10^{-10}$
& $2\times 10^{-8}$
& $2\times 10^{-7}$
& $2\times 10^{-6}$
      \\
& $4$
& $2\times 10^{-21}$
& $3\times 10^{-16}$
& $4\times 10^{-13}$
& $5\times 10^{-11}$
& $1\times 10^{-9}$
& $2\times 10^{-8}$
      \\
& $5$
& $2\times 10^{-26}$
& $5\times 10^{-20}$
& $3\times 10^{-16}$
& $1\times 10^{-13}$
& $9\times 10^{-12}$
& $2\times 10^{-10}$
      \\
\hline
\hline
\end{tabular*}
\end{table*}

There are also interesting implications for neutrino masses \cite{numass}. A very
light
(non-seesaw) doublet neutrino Majorana mass is possible from the superpotential
\begin{equation}
\label{wmaj1}
W\sim\frac{H_2\cdot
L_i}{M}\left(\frac{S}{M}\right)^{P^{''}_{L_iL_i}},
\end{equation}
but the generated mass is too small ($< 10^{-4}\ eV$) to be relevant for
dark matter or MSW conversions in the Sun.
If right-handed neutrino chiral multiplets, $\nu_i^c$, are introduced in the
model, then Majorana and Dirac neutrino mass terms can be generated via
\begin{eqnarray}
W_D&\sim & H_2\cdot L_i \nu_i^c \left(\frac{S}{M}\right)^{P^{'}_{L_i
{\nu}^c_i}},\nonumber\\
W_M&\sim & \nu_i^c \nu_i^c S
\left(\frac{S}{M}\right)^{\bar{P}_{{\nu}^c_i{\nu}^c_i}}.
\end{eqnarray}
The superpotential $W_D$ can yield Dirac neutrino masses heavier than those
coming from Eq.~(\ref{wmaj1}). For example, taking $K=5$ and ${P^{'}_{L_i
{\nu}^c_i}}=4$ or 5 gives masses $m_{L_i {\nu}^c_i} = 0.9$ eV or $10^{-2}$ eV,
respectively, which are in the interesting range for solar and atmospheric
neutrinos, oscillation experiments and dark matter.

The superpotential $W_M$, on the other hand, can give a Majorana mass to the
singlet neutrinos $\nu_i^c$ which can be  very large or small, depending on the
sign of $\bar P_{{\nu}^c_i {\nu}^c_i}-K$.
Combining the effects of $W_M$ and $W_D$, light neutrinos can be produced by the
standard seesaw mechanism, with mass
\begin{eqnarray}
m^{\rm light}_{\rm seesaw} &\sim &
m_{L_i {\nu}^c_i}^2/ m_{{\nu}^c_i {\nu}^c_i}\nonumber\\
& \sim & m_{soft}
\left(\frac{m_{soft}}{M}\right)^{\frac{2P'_{L_i {\nu}^c_i}
+ K - \bar{P}_{{\nu}^c_i {\nu}^c_i}}{K+1}}.
\end{eqnarray}
With the choice $K=5$ and $P'_{L_i {\nu}^c_i}= P'_i = \{2,1\}$
for $i=1,\, 2$, respectively and with either $P'_{L_3{\nu}^c_3} = 1$
or $P'_{L_3{\nu}^c}= P'_{u_3} = 0$
(involving a renormalizable Dirac neutrino term) the light
eigenvalues of the three generations fall into the hierarchy of
$3\times 10^{-5}$ eV, $1\times 10^{-2}$ eV, and either $1\times 10^{-2}$ eV 
or $5$ eV for $\bar{P}_{{\nu}^c_i {\nu}^c_i}= P'_{L_i {\nu}^c_i} + 1$.
This range is again of interest for laboratory and non-accelerator experiments.

\section{SUMMARY AND CONCLUSIONS}

A large class of potentially realistic string models predict a gauge group of the
observable sector with additional abelian factors besides the standard group. In
this talk I have discussed some phenomenological and theoretical implications of
this type of models concentrating for simplicity in the simplest case, with just
one additional (non-anomalous) $U(1)'$. This modest extension already has a 
plethora of interesting consequences, some of which I have described. The goal has
been to explore the generic features of these models rather than to construct a
realistic string model. In this respect one can divide the scenarios in two
different classes according to the scale at which the $U(1)'$ symmetry gets
broken.

Generically, $U(1)'$ breaking takes place below the TeV scale much in the same
way
that electroweak breaking occurs. Actually, when the two MSSM Higgs doublets that
break $SU(2)_L\times U(1)_Y$ are charged under $U(1)'$ (which is the case of
interest) the breaking of $U(1)'$ is linked to electroweak breaking. To avoid
fine-tuning the parameters of the model, the scale of $U(1)'$ breaking cannot be
much larger than the Fermi scale. Thus, the new $Z'$ boson is expected to be on
the reach of the next generation of colliders.
An acceptable symmetry breaking pattern requires the presence of at least one 
standard model singlet $S$ charged under the $U(1)'$ and taking a VEV. If the
coupling $SH_1\cdot H_2$ is present in the superpotential, the $\mu$ parameter
is generated dynamically and with the correct order of magnitude.

In this setting, two main possibilities exist for a phenomenologically viable
scenario: a) symmetry breaking driven by the trilinear soft term associated to 
$SH_1\cdot H_2$ and b) radiative breaking with a singlet VEV large  compared to
the electroweak scale (by say, one order of magnitude). Both cases can give 
an acceptably small $Z-Z'$ mixing angle and evade collider bounds on $M_{Z'}$.
Case a) corresponds to a relatively light $Z'$ while case b) gives a heavy
$Z'$. With minimal particle content (no exotics), both scenarios can arise as the
low energy limit of a string model, if some degree of non-universality in the soft
masses is present at the string scale. In the presence of exotics, the large $s$
scenario can be obtained even keeping universal boundary conditions.
Besides solving the $\mu$ problem, these scenarios have interesting implications
for Higgs physics, the mass spectrum of superpartners and dark matter, among
others.

An exception to the generic case described above can occur if the $U(1)'$ is
broken along a flat direction. Such flat directions exists if at least
two SM singlets have $U(1)'$ charges of opposite signs ($F$-flatness is also
required). In that case, the field along the flat direction, $S$, can take a very
large VEV. This is what happens if the potential $V(S)$ gets destabilized at the
origin by a negative soft mass squared $m_S^2$ for $S$ in the infrared. Two main
classes of scenarios are possible depending on the mechanism that stabilizes the
potential at large $S$ and fixes the VEV [and thus the scale of $U(1)'$ breaking]:
in one, the potential is stabilized by $m_S^2$ running to positive values in the
ultraviolet; in the second, stabilization is due to nonrenormalizable operators.
This latter case is more predictive as the scale
involved is fixed mainly by the (discrete) order of the NROs. In both cases the
low-energy
effective theory is simply the MSSM but many of its parameters are
influenced by the intermediate scale breaking of the $U(1)'$. There are, for
example, effects on the soft masses for superpartners, which receive new $U(1)'$
contributions. In addition, NROs coupling MSSM fields to the flat direction can
offer an explanation for the origin of the $\mu$ parameter, as well as for the
hierarchy of the observed fermion masses (and there are also interesting
implications for neutrino masses).

In conclusion, it is worth pursuing the detailed study of such extensions of the
minimal supersymmetric standard model. These simple models are not only well
motivated from the string perspective but have a remarkably large number of
potentially interesting implications, both phenomenologically and theoretically.

\section*{Acknoledgments}
I thank G. Cleaver, M. Cveti\v c, D.A. Demir, B. de Carlos, L. Everett and P.
Langacker for very enjoyable collaborations on the topics covered.

\end{document}